\documentclass{amsart}
\usepackage{amsmath}
\usepackage{amssymb}
\usepackage{amsfonts}

\newtheorem{theorem}{Theorem}

\newtheorem{claim}[theorem]{Claim}

\newtheorem{corollary}[theorem]{Corollary}

\newtheorem{definition}[theorem]{Definition}
\newtheorem{example}[theorem]{Example}

\newtheorem{lemma}[theorem]{Lemma}

\newtheorem{proposition}[theorem]{Proposition}
\newtheorem{remark}[theorem]{Remark}

\newcommand{\cA}{{\mathcal A}}
\newcommand{\cB}{{\mathcal B}}

\newcommand{\gD}{{\mathfrak D}}
\newcommand{\cK}{{\mathcal K}}
\newcommand{\cL}{{\mathcal L}}
\newcommand{\cS}{{\mathcal S}}
\newcommand{\cH}{{\mathcal H}}
\newcommand{\ft}{{\mathbf f}}

\newcommand{\bR}{{\mathbb{R}}}
\newcommand{\bC}{{\mathbb{C}}}

\newcommand{\Cs}{{{$\hbox{\bf C}^*$}}}
\DeclareMathOperator{\Exp}{Exp}
\DeclareMathOperator{\Ext}{Ext}
\DeclareMathOperator{\Tr}{Tr}
\begin{document}

\title{On the structure of positive maps; finite dimensional case}

\author{W{\l}adys{\l}aw A. Majewski}
\address{Institute of Theoretical Physics and Astrophysics, Gda{\'n}sk
University, Wita Stwo\-sza~57, 80-952 Gda{\'n}sk, Poland} 
\email{fizwam@univ.gda.pl}

\begin{abstract}
A natural and intrinsic characterization of the structure of the set $\mathfrak{C}$ of positive unital  maps is given, i.e.  it is shown that $\mathfrak{C}$ is isometrically isomorphic to the subset $\gD$ of bp-positive density matrices endowed with the geometry given by the norm $\alpha$ dual to the Grothendieck projective norm $\pi$, the structure of $\gD$ is determined by the set of its exposed points, and finally a characterization of exposed points of $\gD$ in terms of convex analysis is presented.
This seems to be an answer to an old open problem, characterization of the structure of the set of positive maps,  studied both in Quantum Information and Operator Algebras.
Our arguments are based on the concept of exposed points and links between tensor products and mapping spaces. 
It should be emphasized that the concept of exposed point depends strongly on the geometry, hence the role of the norm $\alpha$ is crucial.
\end{abstract}
\maketitle
\section{Preliminaries}

In this section we summarize some basic facts on the theory of positive maps on the ordered structures with an emphasis on convex structures.
To begin with, let $\cA$ and $\cB$ be \Cs-algebras (with a unit), $\cA_h = \{ a \in \cA:
a = a^* \}$ -- the set of all self adjoint elements in $\cA$, $\cA^+ = \{ a \in \cA_h: a \ge 0 \}$ -- the set of all positive elements in 
$\cA$, and $\cS(\cA)$ the set of all states on $\cA$, i.e. the set of all linear functionals $\varphi$
on $\cA$ such that $\varphi(1) = 1$ and $\varphi(a)\geq0$ for any $a \in \cA^+$.
In particular
$$ (\cA_h, \cA^+)\text{ is an ordered  Banach  space.} $$
We say that a linear map $\alpha : \cA \to \cB$ is positive if $\alpha(\cA^+) \subset \cB^+$.
The set of all linear, bounded (unital) positive maps $\alpha: \cA \to \cB$ will be denoted by $\cL^+(\cA, \cB)$ ($\cL^+_1(\cA, \cB)$ respectively).
Clearly, the set $\cL^+(\cA, \cB)$ is a convex set, i.e. the line segment
$$ [\alpha, \alpha^{\prime}] = \{ \lambda \alpha +(1 - \lambda) \alpha^{\prime}; \quad 
0\leq \lambda\leq1 \}$$
is entirely contained in $\cL^+(\cA, \cB)$ whenever its endpoints $\alpha$ and $\alpha^{\prime}$ are in 
$\cL^+(\cA, \cB)$. We say that $\alpha \in \cL^+_1(\cA, \cB)$ is an extreme point of $\cL^+_1(\cA, \cB)$
if there are no two different maps $\alpha_1$ and $\alpha_2$ in $\cL^+_1(\cA, \cB)$ such that 
$\alpha = \lambda \alpha_1 + (1 - \lambda) \alpha_2$ with $\lambda \in (0,1)$.
The set of extreme points of $\cL^+_1(\cA, \cB)$ will be denoted by $\operatorname{Ext} \{\cL^+_1(\cA, \cB)\}$.
We recall (Krein-Milman theorem) that a compact convex set $C$ is a (closed) convex hull of its extreme points.
In this context it is worth pointing out that if $\cB$ is a von Neumann algebra, then the set of maps of norm $1$ in $\cL^+(\cA, \cB)$
is compact in the point-open topology (see \cite{KadPAMS}). Hence, the study of extremal positive maps is a natural consequence of the Krein-Milman theorem. The following maps, frequently used in quantum theories,  provide a nice illustration of positive extremal maps (see \cite{StLNP})
\begin{enumerate}
\item Jordan homomorphisms $\alpha : \cA \to \cB$,
\item Let $\cA\subset \cB(\cK)$, $\cB \subset \cB(\cH)$. Define a map $\alpha: \cA \to \cB $ to be of the form
$\alpha(A) = V^* A V$, where $V$ is a linear isometry of the Hilbert space $\cH$ into $\cK$.
\end{enumerate}

To sum up this part, there is a natural problem to find all extremal maps in $\cL^+(\cA, \cB)$. In the very special case, for unital maps if $\cA \equiv M_2(\bC)$
and $\cB \equiv M_2(\bC)$, the extremal maps were specified in \cite{StActa}. Some partial results in this direction, for the case $M_2(\bC) \to M_{n+1}(\bC)$ with $n\geq 2$, were obtained in \cite{MajMar} while more generally, positive maps on low dimensional matrix algebras were studied in \cite{W}. However, taking into account the St{\o}rmer remark (see \cite{StLNP})
{\it ``the result {\rm (in \cite{StActa})} being so complicated that it seems to be unfruitful to try to do the same for more general  \Cs algebras''} one may  suppose that the program of finding all extremal positive maps can be too difficult (cf also \cite{GS}). Therefore, we turn to a special subset of extremal positive maps.

\begin{definition}
Let $C$ be a convex set in a Banach space $X$. A point $x \in C$ is an exposed point of $C$ ($x \in \Exp\{C\}$) 
if there is $f \in X^*$ (dual of $X$) such that $f$ attains its maximum on $C$ at $x$ and only at $x$.
\end{definition}

In other words, we wish to have $\langle f,x \rangle > \langle f,y \rangle$ for $x\neq y$ (where $\langle f, x \rangle \equiv f(x)$).  
Clearly, this reflects a kind of a variational principle, both in a mathematical sense (see \cite{Phelps}) as well as in the standard physical sense.
In general, one has $\Ext\{C\} \supseteq \Exp\{C\}$ but there are simple examples of 2-dimensional convex compact sets such that the inclusion $\Ext\{C\} \supset \Exp\{C\}$ is proper (see \cite{FLP}).

Our interest in exposed points stems from the following result (see \cite{Strasz}, \cite{Klee} and \cite{FLP})

\begin{proposition}
Every norm-compact convex set $C$ in a Banach space $X$ is the closed convex hull of its exposed points.
\end{proposition}

Definition of exposed point is a specification of more general concept (see \cite{ASI}, pp 1-4)

\begin{definition}
\begin{enumerate}

\item A convex subset F of a convex set $K\subset X$ is defined to be a face of K if the condition $\omega \in F$, $\omega = \sum_i \lambda_i \omega_i$, $\lambda_i \geq 0$, $\sum_i \lambda_i = 1$, and $\omega_i \in K$ imply $\omega_i \in F$.
\item F is exposed if there exists $f \in X^*$ and an $\alpha \in \bR$, such that $f(x) = \alpha$ for all $x \in F$ and $f(y) < \alpha$ for all $y \in K \setminus F$.
\end{enumerate}
\end{definition}

In particular, an exposed point is a $0$-dimensional exposed face. On the other hand, we will see, in Section 2,  that our basic object $\mathfrak{D}$
(describing the set of all  positive, unital maps) will be an exposed face.

To speak about exposed points for $\cL^+(\cA, \cB)$ we should describe 
 the corresponding linear duality in an effective way. To this end, firstly, we note that
in his pioneering work on Banach spaces, Grothendieck \cite{Gro} observed the links between tensor products and mapping spaces. To describe these, we will select certain results from the theory of tensor products of Banach spaces. The point is that the  ``union'' of the linear structure of tensor products with a topology is not unique - namely, there are many ``good'' cross-norms (cf \cite{Tak}; pp 188-189, p. 229). However, among them, there is the projective norm which gives rise to the projective tensor product and this tensor product linearizes bounded bilinear mappings just as the algebraic tensor product linearizes bilinear mappings (see \cite{Ryan}; p. 22).

Let $X$, $Y$ be Banach algebras. 
We denote by $X \odot Y$ the algebraic tensor product of $X$ and $Y$
 (algebraic tensor product of two $^*$-Banach algebras is defined as tensor product of two vector spaces with $^*$-algebraic structure determined by the two factors; so the topological questions are not considered). 
We consider the following (projective) norm on $X \odot Y$
\begin{equation}
\label{projektywna norma}
\pi(u) = \inf \{ \sum_{i=1}^{n}\Vert x_i \Vert \Vert y_i\Vert: \quad u = \sum_{i=1}^n x_i \otimes y_i, \quad x_i \in X, \quad y_i \in Y\}.
\end{equation}

We denote by $X \otimes_{\pi}Y$ the completion of $X \odot Y$ with respect to the projective norm $\pi$ and this Banach space will be referred to as the projective tensor product of the Banach spaces $X$ and $Y$.
 
Denote by $\mathfrak{B}(X \times Y)$ the Banach space of bounded bilinear mappings $B$ from $X \times Y$ into the field of scalars with the norm given by $||B|| = \sup \{ |B(x,y)|; \Vert x \Vert \leq 1, \Vert y \Vert \leq 1 \}$.
Note (for all details see \cite{Ryan}; Section 2.2), that with each bounded bilinear form $B \in \mathfrak{B}(X \times Y)$ there is an associated operator $L_B \in \cL(X, Y^*)$ defined by $\langle y, L_B(x)\rangle = B(x,y)$.  The mapping $B \mapsto L_B$ is an isometric isomorphism between the spaces $\mathfrak{B}(X \times Y)$ and $\cL(X, Y^*)$. Hence, there is an identification
\begin{equation}
\label{lala}
(X \otimes_{\pi} Y)^* = \cL(X, Y^*),
\end{equation}
such that the action of an operator $S:X \to Y^*$ as a linear functional on $X \otimes_{\pi}Y$ is given by
\begin{equation}
\label{haha}
\langle \sum_{i=1}^n x_i \otimes y_i , S \rangle = \sum_{i=1}^n \langle y_i, Sx_i \rangle.
\end{equation}

Note that identification (\ref{lala}) and  relation (\ref{haha}) give the solution for the description of linear duality
which is required for the definition of exposed points of $\cL^+(\cA, \cB)$.
This will be the starting point in the characterization of exposed positive maps and this will be done in the next Sections. We wish to complete the presented material by
 recalling  another well known result (see \cite{St1}) which will be crucial in our work. Moreover, it can serve as an illustration to the given material as well as to indicate that the relation (\ref{lala}) is very relevant to an analysis of positive maps. To present the above mentioned result we need some preparations. 

Let $\mathfrak{A}$ be a norm closed self-adjoint subspace of bounded operators on a Hilbert space $\mathcal K$ containing identity operator. $\mathfrak T$ will denote the set of trace class operators on a Hilbert space $\cH$. $\cB(\cH) \ni x \mapsto x^t \in \cB(\cH)$ stands for the transpose map of $\cB(\cH)$ with respect to some orthonormal basis.
The set of all linear bounded (positive) maps $\phi: \mathfrak{A} \to \cB(\cH)$ will be denoted by $\cL(\mathfrak{A}, \cB(\cH))$ ($\cL^+(\mathfrak{A}, \cB(\cH))$ respectively). 
 Finally, we denote by ${\mathfrak A} \odot {\mathfrak T}$ the algebraic tensor product of $\mathfrak A$ and $\mathfrak T$
  and  ${\mathfrak A} {\otimes}_{\pi} \mathfrak T$ means its Banach space closure under the projective norm defined similarly as in (\ref{projektywna norma}) 
  
\begin{equation}
\label{4}
\pi(x) = \inf \{ \sum_{i=1}^n \Vert a_i \Vert  \Vert b_i \Vert_1: x = \sum_{i=1}^n a_i \otimes b_i, \ a_i \in {\mathfrak A}, \ b_i \in {\mathfrak T} \},
\end{equation}
where $\Vert \cdot \Vert_1$ stands for the trace norm. Now, we can quote (see \cite{St1})
\begin{lemma} (Basic)
\label{pierwszy lemat}
There is an isometric isomorphism $\phi \mapsto \tilde{\phi}$ between $\cL({\mathfrak A}, \cB(\cH))$ and
$({\mathfrak A} {\otimes}_{\pi} {\mathfrak T})^*$ given by
\begin{equation}
\label{5}
(\tilde{\phi})(\sum_{i=1}^n a_i\otimes b_i) = \sum_{i=1}^n \Tr(\phi(a_i)b^t_i),
\end{equation}
where $\sum_{i=1}^n a_i\otimes b_i \in {\mathfrak A}\odot {\mathfrak T}$.

Furthermore, $ \phi \in \cL^+({\mathfrak A}, \cB(\cH))$ if and only if $\tilde{\phi}$ is positive on ${\mathfrak A}^+ {\otimes}_{\pi} {\mathfrak T}^+$.
\end{lemma}

To comment on this result we make
\begin{remark}
\begin{enumerate}
\item $\mathfrak T$ appears in Lemma \ref{pierwszy lemat} as ${\mathfrak T}^* = \cB(\cH)$ (so in expression (\ref{lala}), one puts $Y = \mathfrak T$).
\item There is no restriction on the dimension of Hilbert spaces. In other words, this result can be applied to the true quantum systems.
\item In \cite{St2}, St{\o}rmer showed that in the special case when ${\mathfrak A} = M_n(\bC)$ and $\cH$ has dimension equal to $n$, the above Lemma is a reformulation of Choi result \cite{Ch1}, \cite{Ch3}.
\item  One should note that the positivity  of the functional  $\tilde{\phi}$ is defined by the cone
${\mathfrak A}^+ {\otimes}_{\pi} {\mathfrak T}^+$ (for another definitions of positivity in tensor products see \cite{Witt}, and \cite{MMO}). In particular, positivity determined by $({\mathfrak A} \odot {\mathfrak T})^+$ leads to completely positive maps (cf. \cite{Tak}, \cite{I}, and \cite{MMO}).
\item Note that definition of $\pi(\cdot)$ leads to the proper isometricity although it gives some problems with normalization (e.g. see definition of $\mathfrak{D}$ in the next Section).
\item Some forms of Basic Lemma, being a consequence of Grothendieck's approach, are known at least from late sixties, see \cite{WW} pp 45-46 and references given there.
\end{enumerate}
\end{remark}

\vskip 0.5cm

The aim of this paper is to give a  characterization of the structure of positive, unital maps. This will be done for a finite dimensional case. 
However, it should be emphasized that the used strategy, concepts and basic facts can be easily generalized. On the other hand, finite dimensionality makes our work more accessible for people working in Quantum Information.

A description of exposed positive maps is the main difficulty in carrying out the sought characterization. To this end, in Section 2, 
 we will give a characterization of normalized block positive density matrices corresponding to unital positive maps. Section 3 provides illustrative examples.  Then, in Section 4, an analysis of exposed points will be provided.
 The identification (\ref{5}) (so, the specification of (\ref{lala})) between positive maps and block positive density matrices will be our basic tool.  Note, that from the methodological point of view, our strategy can be considered as an extension and generalization of the discussion given in Section X of \cite{MMO} combined with the construction dual to the Choi approach (cf \cite{majphd}
 and \cite{LMM}).

We wish to close this Section with one more definition (see \cite{BDM} and \cite{terdal}). 
\begin{definition}
\label{defunex}
Let $\cH = \cH_1 \otimes \cH_2$ be a finite dimensional Hilbert space. A partial product basis is a set of mutually orthonormal simple tensors spanning a proper subspace $\cH_0$ of $\cH$. An unextendible product basis is a partial product basis whose complementary subspace $\cH^{\bot}_0$ contains no simple tensor.
\end{definition}

This concept will clarify some of geometrical aspects of certain exposed block positive density matrices, see next Sections.

\section{Normalized block positive density matrices}

Let us specify Basic Lemma for finite dimensional case with slight modifications. We put ${\mathfrak A} = \cB(\cH)$ with $\cH$ being a finite dimensional Hilbert space. In this case, ${\mathfrak T} = \cB(\cH)$ and we arrive at the following isomorphism
$\phi \mapsto \tilde{\phi}$ between $\cL(\cB(\cH), \cB(\cH)) \equiv \cL( \cB(\cH))$ and $(\cB(\cH) \otimes_{\pi} \cB(\cH))^*$
such that $\phi \in \cL^+(\cB(\cH))$ if and only if $\tilde{\phi}$ is positive on $\cB(\cH)^+ \otimes_{\pi} \cB(\cH)^+$ (here and subsequently, $(\cB(\cH) \otimes_{\pi} \cB(\cH))$ denotes the projective tensor product with the norm $\pi$ given by (\ref{4})).
Further, note that identifying the real algebraic tensor product $\cB(\cH)_h \odot \cB(\cH)_h$ of self-adjoint parts of $\cB(\cH)$ with a real subspace of $\cB(\cH) \odot \cB(\cH)$, one has $\cB(\cH)_h \odot \cB(\cH)_h = (\cB(\cH) \odot \cB(\cH))_h$. Obviously, this can be extended for the corresponding closures.  From now on, we will use these identifications and we will study certain subsets of real tensor product spaces.

The next easy observation says that the discussed isomorphism sends the set $\Exp \{ \cL^+(\cB(\cH)) \}$ onto the set $\Exp \{(\cB(\cH) \otimes_{\pi} \cB(\cH))^{*,+} \} $, where $(\cB(\cH) \otimes_{\pi} \cB(\cH))^{*,+}$ stands for functionals 
on $\cB(\cH) \otimes_{\pi} \cB(\cH)$
which are positive on $\cB(\cH)^+ \otimes_{\pi} \cB(\cH))^+$.

Therefore, our task can be reduced to a study of exposed points of the last set.
Let us elaborate upon this point. Any (linear, bounded) functional in $( \cB(\cH) \otimes_{\pi} \cB(\cH))^{*,+}$ is of the form
\begin{equation}
\label{5a}
\varphi(x \otimes y) = \Tr \varrho_{\varphi} \ x \otimes y,
\end{equation}
(the finite dimension case is assumed; so any functional is automatically normal) with $\varrho_{\varphi}$ being a ``density'' matrix satisfying the following positivity condition (frequently called ``block-positivity'', and denoted ``bp'' for short)
\begin{equation}
\varrho_{\varphi} \geq_{bp}0 \quad  \Leftrightarrow  \quad (f \otimes g, \varrho_{\varphi} f \otimes g) \geq 0,
\end{equation}

for any $f,g \in \cH$.
To take into account that the isomorphism given in Lemma 3 is also isometric, note that  $\cL(\cB(\cH))$ is equipped with the Banach space operator norm $\Vert \cdot \Vert$. 
On the other hand,  formula (\ref{4}) defines the cross - norm, which is not smaller than max \Cs-norm. 

\begin{definition}
The dual norm $\alpha$ to the projective norm $\pi$ is defined as
\begin{equation}
\alpha(\varrho_{\varphi}) = \sup_{0\neq a \in 
\cB(\cH) \otimes_{\pi} \cB(\cH)} \frac{|\Tr \varrho_{\varphi} a|}{\pi(a)}.
\end{equation}
\end{definition}

An application of Proposition IV.2.2 in \cite{Tak} (see also \cite{Schat}) shows that $\alpha(\cdot)$ is well defined cross-norm.
It will be useful to note that for bp density matrix $\varrho_{\varphi}$
\begin{equation}
\alpha(\varrho_{\varphi}) \geq \frac{|\Tr \varrho_{\varphi}|}{\pi(I)} \quad \text{and} \quad  |\Tr\varrho_{\varphi}| = \Tr \varrho_{\varphi}\leq n \alpha(\varrho_{\varphi}),
\end{equation}
where we have used that $\pi(I) = n$ and $I$ stands for the unit.

Define
\begin{equation}
\label{Do}
{\mathfrak D}_0 = \{ \varrho_{\varphi}: \alpha(\varrho_{\varphi}) = 1, \  \varrho_{\varphi}= \varrho_{\varphi}^*, \ \varrho_{\varphi}\geq_{bp} 0 \}.
\end{equation}

Basic Lemma and the above discussion say that there is an isometric isomorphism between the set of positive linear maps in  $\cL(\cB(\cH))^+$ of norm one and ${\mathfrak D}_0$. Moreover, for each $\varrho_{\varphi} \in {\mathfrak D}_0$ one has $\Tr \varrho_{\varphi} \leq n \alpha(\varrho_{\varphi}) = n$.

To proceed with the analysis of Basic Lemma we note that the formula (\ref{5}) 
says that any bp density matrix $\varrho_{\phi}$ determined by an unital positive map $\phi$ has the following normalization

\begin{equation}
\label{8}
\Tr \varrho_{\phi} \equiv \Tr_{\cH \otimes \cH} \varrho_{\phi} = \Tr_{\cH} \phi(I) = n. 
\end{equation}

Conversely, assume that $\varrho_{\phi} \in {\mathfrak D}_0$ and $\Tr \varrho_{\phi} = n$. Then, there exists a linear positive bounded map $\phi$ (of norm one) such that  $\Tr \phi(I)I = n$.

However,
\begin{equation}
\Vert \phi(I) \Vert \leq \Vert \phi \Vert \Vert I \Vert = \alpha(\varrho_{\phi}) \cdot 1 = 1. 
\end{equation}
Thus $\Vert \phi(I)\Vert \leq 1$. But $\phi(I) \geq 0$ has the following spectral decomposition
$$ \phi(I) = \sum_i \lambda_i E_i ,$$
with $\lambda_i \leq 1$ for all i, and $\sum_i \lambda_i = n$. This means that $\phi(I) = I$.

To sum up, firstly, we give

\begin{definition}
The set of bp normalized density matrices is defined as
\begin{equation}
\mathfrak{D} = \{ \varrho_{\phi}: \alpha(\varrho_{\phi}) = 1, \  \varrho_{\phi}= \varrho_{\phi}^*, \ \varrho_{\phi}\geq_{bp} 0, \ \Tr\varrho_{\phi} = n \},
\end{equation}
\end{definition}

then we have 
\begin{corollary}
 Lemma \ref{pierwszy lemat} gives an isometric isomorphism between the convex set of unital positive maps $\mathfrak{C} \equiv \cL^+_1(\cB(\cH))$ and the subset $\mathfrak{D}$ 
of bp normalized density matrices. \end{corollary}.

\begin{remark}
\label{dobrauwaga}
\begin{enumerate}
\item
Geometrically speaking, we are using the correspondence between two ``flat'' subsets of  balls in $\cL(\cB(\cH))$ and in the set of all self-adjoint density matrices $(\cB(\cH) \otimes_{\pi} \cB(\cH))_h$, respectively. The interest of this remark follows from the fact that the considered  balls are not so nicely shaped as the closed  ball of real Euclidean 2-- or 3 dimensional space (see Chapter 5 in \cite{Megg} for geometrical details).
\item
One can easily show that $\mathfrak{D}$ is an exposed face in $\mathfrak{B}_1^{(+)} \equiv \{  \varrho_{\varphi}: \alpha(\varrho_{\varphi}) \leq 1, \  \varrho_{\varphi}= \varrho_{\varphi}^*, \ \varrho_{\varphi}\geq_{bp} 0 \}$. The principal significance of this remark is in the so called ``transmission of extremality''. Namely, (see \cite{HUL},  Sections 2.3 and 2.4),
if $x \in \mathfrak{D}$ is an extreme point of $\mathfrak{B}_1^{(+)}$, then it is a fortiori an extreme point of $\mathfrak{D}$. But, as $\mathfrak{D}$ is a face of $\mathfrak{B}_1^{(+)}$, the converse is also true.
The discussed property of transition of extremality also applies to exposed faces. Due to this fact the analysis of extreme points is easier, at least for some special cases.
 Finally, we remark that there is no a straightforward generalization of that property for exposed points (see \cite{Rock}, pp 162-163).
\end{enumerate}
\end{remark}

We wish to look more closely at the structure and properties of $\mathfrak D$. To state the first result we need:
\begin{definition}
Let $f \in \cH \otimes \cH$. The Schmidt decomposition (see \cite{EK}) is given as
\begin{equation}
\label{Schmidt}
f = \sum_i^{N_s} g_i u_i \otimes v_i,
\end{equation}
with  two sets of mutually orthogonal vectors $\{u_i \}_i$ and $\{ v_i \}_i$ in $\cH$. $g_i \in \bC$, and $\sum_i |g_i|^2 = \Vert f \Vert^2$. $N_s$, the number of nonzero coefficients in (\ref{Schmidt}), is called the Schmidt rank of $f$.
\end{definition}

\begin{lemma}
\label{fajnylemat}
Let $f$ be in $\cH\otimes \cH$, and $\Vert f \Vert = 1$.
Then $1\leq \pi(|f><f|) \leq N_s, $ where $|f><f|g \equiv (f,g)f$ for any $g \in \cH$.
\end{lemma}
\begin{proof}
As the projective norm $\pi$ is submultiplicative (see \cite{Tak}, p. 205) or it can be easily and directly checked) one has
$\pi(|f><f|) \leq (\pi(|f><f|))^2$, so 
\begin{equation}
\label{Ohh}
1 \leq \pi(|f><f|).
\end{equation}
On the other hand, employing the Schmidt decomposition 
and using another formula for computation of the projective norm $\pi$ (see Section 2.2 in \cite{Ryan}) one has
\begin{equation}
\label{.}
\pi(|f><f|) = \sup \{ |\sum_{ij}g_ig^*_j \Tr |u_i><u_j| S (|v_i><v_j|)|: S \in \cL(\cB(\cH), \cB(\cH)), \Vert S\Vert\leq 1 \}
\end{equation}
$$= \sup_S  \{ |\sum_{ij} g_i g^*_j \sum_k (u_k,u_i)(u_j, S(|v_i><v_j|) u_k)|\}$$
$$= \sup_S \{ | \sum_{ij} g_i g^*_j (u_j, S(|v_i><v_j|) u_i)|\}$$
$$= \sup_S \{ | \sum_{ij}( \tilde{u_j}, S(|v_i><v_j|) \tilde{u_i})| \} \quad \text{where} \quad  \tilde{u_i}= g_iu_i$$
$$ \leq \sup_S \{ \sum_{ij} |( \tilde{u_j}, S(|v_i><v_j|) \tilde{u_i})| \}$$
$$ \leq \sup_S \{ \sum_{ij} \Vert  \tilde{u_j} \Vert \Vert S \Vert \Vert |v_i><v_j \Vert \Vert  \tilde{u_i} \Vert \} $$
$$ \leq \sum_{ij}\Vert  \tilde{u_i} \Vert \Vert  \tilde{u_j} \Vert \leq \frac{1}{2} \sum_{ij} (\Vert  \tilde{u_i}\Vert^2 + \Vert  \tilde{u_j}\Vert^2) = \frac{1}{2} N_s(\Vert f \Vert^2 + \Vert f \Vert^2) = N_s.$$
Consequently
\begin{equation}
\label{Oh}
1\leq \pi(|f><f|) \leq N_s.
\end{equation}

\end{proof}

The following Lemma yields a useful information about $\mathfrak{D}$:
\begin{corollary}
\label{wniosek1}
$\mathfrak{D}$ is contained in the ball of $(\cB(\cH) \otimes \cB(\cH))_h$  of radius $n$ (with respect to the operator norm.)
\end{corollary}
\begin{proof}
Note that
$$
1 = \alpha(\varrho_{\varphi}) = \sup_{0\neq a \in 
\cB(\cH) \otimes_{\pi} \cB(\cH)} \frac{|\Tr \varrho_{\varphi} a|}{\pi(a)} $$
\begin{equation}
\geq \frac{|\Tr \varrho_{\varphi} |f_i><f_i|}{\pi(|f_i><f_i|)} = \frac{|\lambda_i|}{\pi(|f_i><f_i|)},
\end{equation}
where we have used the spectral decomposition of $\varrho_{\varphi}$
\begin{equation}
\varrho_{\varphi} = \sum_i \lambda_i |f_i><f_i|.
\end{equation}
As $N_s \leq n$, the claim follows.
\end{proof}

Other useful properties of $\mathfrak D$ are collected in the following statements.
\begin{proposition}
\label{LO}
$\mathfrak D$ is globally invariant with respect to the following operations:
\begin{enumerate}
\item local operations, LO for short, i.e. maps implemented by unitary operators
 $U: \cH \otimes \cH \to \cH \otimes \cH$ of the form $U= U_1 \otimes U_2$ where $U_i : \cH \to \cH$ is unitary, $i = 1,2$;
\item partial transpositions $\tau_p = id_{\cH} \otimes \tau : \cB(\cH)\otimes \cB(\cH) \to \cB(\cH) \otimes \cB(\cH)$
where $\tau$ stands for transposition.
\end{enumerate}
\end{proposition}
\begin{proof}
(1)
Note
$\varrho \geq_{bp} 0$ if and only if for any $x,y \in \cH$
$$\Tr \varrho \cdot P_x \otimes P_y  \geq 0,$$
where $P_x \equiv |x><x|$. But
\begin{equation}
\Tr U_1\otimes U_2 \cdot \varrho \cdot U^*_1 \otimes U^*_2 \cdot P_x \otimes P_y = \Tr \varrho \cdot P_{U_1x} \otimes P_{U_2y} \geq 0.
\end{equation}
Furthermore,
\begin{equation}
\alpha(U \ \varrho \ U^*) = \sup_{a\neq 0} \frac{|\Tr U\ \varrho \ U^* a|}{\pi(a)} = \sup_{a\neq 0} \frac{|\Tr \varrho \ U^* a U|}{\pi(a)},
\end{equation}
and
\begin{equation}
\pi(a) = \inf \{ \sum_i \Vert a_i^1\Vert \Vert a^2_i \Vert_1: a = \sum_i a^1_i \otimes a^2_i, \quad a^1_i, a^2_i \in \cB(\cH) \}
\end{equation}
$$ = \inf \{ \sum_i \Vert U_1 a_i^1U^*_1  \Vert \Vert U_2 a_i^2 U^*_2 \Vert_1: a = \sum_i a_i^1 \otimes a_i^2 \} = \pi(U a U^*),$$

which proves the first claim.

(2) Any $\sigma, a \in \cB(\cH) \otimes \cB(\cH)$ can be written as a finite sum of elementary tensors $\sigma = \sum_i \sigma^1_i \otimes \sigma_i^2$ and
$a = \sum_i a_i^1 \otimes a_i^2$ respectively. Note, that
\begin{equation}
\Tr_{(\cH \otimes \cH)} \tau_p(\sigma) \ \tau_p(a) = \sum_{k,l} \Tr_{(\cH \otimes \cH)} \sigma_k^1 a_l^1 \otimes \tau(\sigma_k^2) \ \tau(a^2_l)
\end{equation}
$$= \sum_{k,l} \Tr_{(\cH)}\{ \sigma_k^1 a_l^1\} \Tr_{(\cH)} \{\tau(\sigma^2_k) \ \tau( a^2_l) \}$$
$$= \sum_{k,l} \Tr_{(\cH)} \{\sigma^1_k \ a^1_l \} \Tr_{(\cH)} \{\sigma_k^2 \ a_l^2 \} = \Tr_{(\cH\otimes \cH)} \{ \sigma  a \}.$$
Thus, bp condition is preserved. Moreover
\begin{equation}
\pi(\tau_p(a)) = \inf \{ \sum_i \Vert a_i^1 \Vert \Vert \tau(a^2_i) \Vert_1: a = \sum_i a_i^1 \otimes a^2_i \}
\end{equation}
$$= \inf \{ \sum_i \Vert a_i^1 \Vert \Vert a_i^2 \Vert_1: a = \sum_i a^1_i \otimes a_i^2 \} = \pi(a).$$
Consequently
\begin{equation}
\alpha(\tau_p(\sigma)) = \alpha(\sigma),
\end{equation}
and the second claim is proved.
\end{proof}

To provide the reader with interesting examples of elements of $\mathfrak D$ we will need (see \cite{ASI}, p. 251)
\begin{definition}
A self-adjoint unitary operator $s$ is called a symmetry, i.e. $s = s^*$ and $s^2 = I$.

A self-adjoint operator $s$ is called a partial symmetry (equivalently called e-symmetry) if $s^2$ is an orthogonal projector $e$.
\end{definition}

It is an easy observation (see \cite{ASI}) that each  e-symmetry can be uniquely decomposed as a difference $s = p - q$ of two orthogonal
 projectors, i.e. $p = \frac{1}{2}(e + s)$, $q = \frac{1}{2} (e - s)$, and $pq = 0 = qp$.

 To compute the norm $\alpha(a)$, at least for some interesting examples, it is tempting to use a relation between 
 $\alpha$ and a positive functional of the form $\phi(\cdot) = \Tr\{ a \cdot\}$ with $a$ being a positive operator.
 However, there is a difficulty coming from the fact that $\cB(\cH) \otimes_{\alpha} \cB(\cH)$ is
 a Banach $^*$-algebra with unit but $\alpha(1) = n \neq 1$. Therefore, we can not use the standard results (see \cite{Dix}, or \cite{Nai}).
 Thus, to proceed, in an effective way, an examination of the structure of $\mathfrak D$ we will provide
 
 \begin{lemma}
 \label{norma alfa1}
 Let $\sigma \in \cB(\cH) \otimes_{\alpha} \cB(\cH)$.
 $$\alpha(\sigma) = \max \{ |\Tr \sigma \cdot s\otimes p \ | : s \in \cS(\cH), p \in Proj^1(\cH) \}$$
where $\cS(\cH)$ denotes the set of all symmetries in $\cB(\cH)_h$ while   $Proj^1(\cH)$ stands for the set $\{ \pm |f><f|: f \in \cH, \Vert f\Vert = 1 \}.$
\end{lemma}
\begin{proof}
By definition
\begin{equation}
\label{norma alfa}
\alpha(\sigma) = \sup \{ |\Tr \sigma \ x |: \pi(x)\leq 1 \}
\end{equation}

 But, the unit ball $B(1, \pi)$ (with respect to the norm $\pi$) is convex, compact (finite dimensional case is assumed), the function 
 $x \mapsto |\Tr \sigma \ x \ |$ is convex, continuous. Therefore, Bauer maximum principle (see Theorem 25.9 in \cite{Cho}; in more algebraic context Lemma 4.1.12 in \cite{BR})  implies that $\sup$ in (\ref{norma alfa})
is attainable and it is equal to the value of the function $x \mapsto |\Tr \sigma \ x \ |$ on an extremal point of $B(1, \pi)$.
On the other hand, note (cf  Chapter 2 in \cite{Ryan}) that $B(1, \pi)$ is the closed convex hull of the set $B_1\otimes B_2$,
where $B_1 \equiv B(1, \Vert \ \Vert)$ is the closed unit ball (with respect to the norm $\Vert \ \Vert$)in $\cB(\cH)$ while 
$B_2 \equiv B(1, \Vert \ \Vert_1)$ is the closed unit ball (with respect to the trace norm $\Vert \ \Vert_1$) in $\cB(\cH)$.

It is well known (see  Chapter 7, Proposition 7.4.6 in \cite{KR} or \cite{sakai}) that any extreme point of the unit ball of a self-adjoint 
part of a $C^*$-algebra is given by a symmetry. Finally, note that $\varrho^*= \varrho \in B(1, \Vert \ \Vert_1)$ means
$\varrho = \sum_i \lambda_i P_i$, where $\lambda_i \in \bR$, $P_i$ denotes one dimensional orthogonal projector, and
\begin{equation}
\Vert \varrho \Vert_1 = \sum _i | \lambda_i \ | \leq 1.
\end{equation}
Consequently, self-adjoint part of the unit ball in $(\cB(\cH), \Vert \ \Vert_1)$ is spanned by the set
 $\{ \pm |f><f|; f \in \cH, \Vert f \Vert =1 \}$
what completes the proof.
\end{proof}

\begin{corollary}
\begin{enumerate}
\item Let $P$ be a projector (on $\cH \otimes \cH$). Then $\alpha(P) = \Vert \Tr_{\cH} P \ \Vert$
where $\Tr_{\cH}$ stands for the partial trace.
\item Let $W$ be a symmetry (on $\cH \otimes \cH$). Then $\alpha(W) = \max \{ \Vert \Tr_{\cH}( W s\otimes 1) \ \Vert : s \in \cS(\cH) \}.$
\end{enumerate}
\end{corollary}
\begin{proof}
(1) 
\begin{equation}
\label{28a}
\sup_{\pi(x)\leq 1} |\Tr P \ x | = max \{ |\Tr P \cdot s\otimes p \ | : s \in \cS(\cH), \ p \in Proj^1(\cH) \}
\end{equation}
$$= \max \{ \sum_i (e_i\otimes g, P s\otimes 1 e_i \otimes g)|: s \in \cS(\cH), \ g \in \cH, \ \Vert g \Vert =1 \}  =
\max_{s, g} |(g, \Tr_{\cH_1} P \ s \otimes 1\ g) |, $$
where $\cH_1 \equiv \cH \otimes \bC$, and $\{ e_i \}$ is a basis in $\cH$.

As, $P \geq 0$, a maximum value problem in the trace class operators (see \cite{Schat}) gives
$$ \sup_{\pi(x)\leq 1} |\Tr P \ x |  = \max_{\Vert g \Vert =1} |(g, \Tr_{\cH_1} P \ g) | = \max_{\Vert g \Vert =1} (g, Tr_{\cH} P \ g) = \Vert \Tr_{\cH} P \ \Vert,$$
and the first claim follows.

(2)  Similarly
\begin{equation}
\sup_{\pi(x)\leq 1} |\Tr W \ x \  | = \max_{s,p} | \Tr W \cdot s \otimes p \ |
\end{equation}
$$= \max_{s, g} |(g, \Tr_{\cH} W \cdot s\otimes 1 g)| = \max_{s} \Vert \Tr_{\cH} W \cdot s \otimes1 \Vert$$
and the proof is complete.
\end{proof}

Now, we are in position to give the promised examples.
\begin{example}
\label{fajnyprzyklad}
\begin{enumerate}
\item Any projector of the form $P = p\otimes I$ where p is a one dimensional projector on $\cH$, $I$ is the identity on $\cH$, is an element of $\gD$.
\item Let $\{ e_i \}$ be a basis in $\cH$. Define $f \in \cH \otimes \cH$ by
\begin{equation}
f = \frac{1}{\sqrt{n}}|e_1\otimes e_1 + e_2\otimes e_2 + ... + e_n \otimes e_n>
\end{equation}
Obviously, $n|f><f| \in {\mathfrak D}$.
\item Define 
\begin{equation}
W = \sum_{i,j} E_{ij} \otimes E_{ji}
\end{equation}
where $E_{ij} \equiv |e_i><e_j|$. $W$ is a bp symmetry with $\Tr W = n$. Moreover, $\tau_p(W) = n|f><f|.$ 
Hence, $ W \in \mathfrak D$.
\end{enumerate}
\end{example}

To appreciate the above examples we make

\begin{remark}
\begin{enumerate}
\item In Quantum Information Theory, $W$ is called the swapping operation.
\item $W g_1 \otimes g_2 = g_2 \otimes g_1$
\item If we apply $W$ to the correspondence (\ref{5}) and take into account (\ref{5a}) we get that $\phi$  is a transposition.
\end{enumerate}
\end{remark}

One of the big ``mysteries'' of the structure of positive maps is the appearance of non-decomposable maps for n$D$ 
(n-dimensional) cases with $n\geq 3.$ 
In our approach this means that, for $nD$, $n\geq 3$, $conv(\mathfrak{D}^+, \tau_p(\mathfrak{D}^+))$ is a proper subset of $\mathfrak{D}$, where 
$\mathfrak{D}^+ = \{ \varrho_{\phi}: \alpha(\varrho_{\phi}) = 1, \  \varrho_{\phi}= \varrho_{\phi}^*, \ \varrho_{\phi}\geq 0, \ \Tr\varrho_{\phi} = n \}$,
and $\tau_p$ stands for the partial transposition (cf \cite{MMO}). The following example shows the geometrical differences (related to symmetries) between $2D$ and $3D$ cases.

\begin{example}
\label{przyklad19}
\begin{enumerate}
\item Assume 2$D$ case and let s be a symmetry in $\mathfrak D$. Then $s = p - q$ and $\Tr s = 2$. As $p + q = I$, then $\Tr(p + q) = 4$. 
Hence $\Tr p = 3$ and $\Tr q = 1$. Consequently, $q$ is one dimensional orthoprojector, so $q = |h><h|$ with 
$h \in \cH \otimes \cH$, $\Vert h \Vert =1.$
Applying bp condition to $s$ we obtain
\begin{equation}
\forall_{g_1, g_2 \in \cH} \quad \Vert g_1 \otimes g_2 \Vert^2 \geq  2 \Vert q \ g_1 \otimes g_2 \Vert^2 = 2 |(h, g_1 \otimes g_2)|^2
\end{equation}
The Schmidt decomposition of $h$ has the form
\begin{equation}
h = \sum_i \lambda_i e_i \otimes f_i
\end{equation}
where $\lambda_i \in \bC$, $\{e_i\}$ and $\{f_i\}$ are orthonormal systems in $\cH$. Normalization of $h$ gives 
$|\lambda_1|^2 + |\lambda_2|^2 = 1$ while bp condition leads to
\begin{equation}
\label{2d}
1\geq 2 |\lambda_i|^2, \qquad i=1,2.
\end{equation}
But, in (\ref{2d}), one can not have strict inequalities (this would be in contradiction with the normalization). Hence, $h$
should be of the form
\begin{equation}
h = \frac{1}{\sqrt{2}} (e^{i\varphi} e_1\otimes f_1 + e^{i \psi} e_2 \otimes f_2)
\end{equation}
with $\varphi, \psi \in [0, 2\pi)$. Consequently, up to the transformation implemented by $U= U_1\otimes U_2$, there 
is a room for a symmetry of the type $W$ only.  It is worth pointing out that this symmetry leads to the transposition.
We end our examination of $\mathfrak D$ for $2$D case with a remark that there is no ``room'' for non-trivial e-symmetry.
\item 3$D$ case.
Let $s$ be a symmetry in $\mathfrak D$, Then $s = p - q$ and $\Tr s = 3$. As $p + q = I$, then $\Tr(p + q) = 9$. 
Hence $\Tr p = 6$ and $\Tr q = 3$. Consequently, $q$ is three dimensional orthoprojector which can be written as  
$q = \sum_{\alpha = 1}^3 |f_{\alpha}><f_{\alpha}|$, $\Vert f_{\alpha} \Vert =1$, $(f_{\alpha},f_{\beta}) = \delta_{\alpha, \beta}$. Let $\{e_i \}_1^3$ be a basis in $\cH$. Then $f_{\alpha} = \sum_{ij} \ft^{\alpha}_{ij} \  e_i \otimes e_j$, where $\ft^{\alpha}_{ij} \in \bC$, and $\sum_{ij} | \ft^{\alpha}_{ij}|^2 =1$.  So, bp condition reads: for any pair of normalized vectors $x,y \in \cH$ one has
\begin{equation}
\label{warunek1}
\Vert x \otimes y \Vert^2 \geq 2 \sum_{\alpha =1}^3 |\sum_{ij} \overline{\ft^{\alpha}_{ij}} (e_i \otimes e_j, x \otimes y)|^2
\end{equation}

Note that the choice : $\ft^{1}_{12} = \frac{1}{\sqrt{2}}, \ \ft^{1}_{21} = - \frac{1}{\sqrt{2}}, \ \ft^{2}_{13} = \frac{1}{\sqrt{2}}, \ \ft^{2}_{31} = - \frac{1}{\sqrt{2}}, \ \ft^{3}_{23} = \frac{1}{\sqrt{2}}, \ \ft^{3}_{32} = - \frac{1}{\sqrt{2}}$ with the other $\ft^{\alpha}_{ij}$ equal to zero, satisfies  condition (\ref{warunek1}) and gives the swapping operation $W$ (clearly for 3$D$ case).

But, $3D$ case offers a new type of normalized bp matrices in $\gD$. Namely, let $x_0 = \frac{1}{\sqrt{2}}(e_1 \otimes e_1 + e_2 \otimes e_1)$.
Denote by $P_v$ an orthogonal projection on the vector $v$ and consider
$$\sigma= P_{x_o} + \sum_{i,j=2}^3 E_{ij} \otimes E_{ij} \equiv P_{x_o} + W_0.$$
$\sigma$, being a hybrid of a projection $P_{x_0}$ and a partial symmetry $W_0$ is an element of $\gD$.
\end{enumerate}
\end{example}

\section{Exposed points of subsets of $\gD$}
To illustrate how the definition of exposed point is working, we will show examples of such points which are in one-to-one correspondence with some basic examples of positive maps.

As the first step we note that normalization of bp matrices enables us to take the functional $f$ appearing in 
the definition of the exposed point (see Definition 1) to be positive.
Namely, let $\varrho_0$ be an exposed point of $\mathfrak{D}$. Then, Definition 1 implies the existence of self adjoint operator $a_f$ such that 
\begin{equation}
\Tr \sigma \cdot  a_f  < \Tr \varrho_0 \cdot a_f
\end{equation}
for any $\sigma \neq \varrho_0$, $\sigma \in \mathfrak{D}$. Let us put
\begin{equation}
a_{f^{\prime}} = a_{f} + \frac{c}{n} \cdot I 
\end{equation}
with a constant $c > 0$, and note that
\begin{equation}
\label{13}
\Tr \sigma \cdot a_{f^{\prime}} = \Tr \sigma \cdot a_{f} + c < \Tr \varrho_0 \cdot a_{f} + c = \Tr \varrho_0 \cdot  a_{f^{\prime}}
\end{equation}
for any $\sigma \neq \varrho_0$, $\sigma \in \mathfrak{D}$.
This  observation and the fact that both sides of (\ref{13}) are linear in $a_f$ (so we can perform the normalization) lead 
to the conclusion that we can always take $f$ to be a positive normalized functional.

The above argument means, 
in functional terms,  that 
an element $\varrho_0$ is an exposed point in $\gD$ if and only if there is a positive operator 
$0\leq a \in
\cB(\cH) \otimes_{\pi} \cB(\cH)$ 
such that

\begin{equation}
\label{A}
\Tr a \cdot \varrho_0 \equiv \langle a,\varrho_0 \rangle  >  \langle a,\sigma\rangle \equiv \Tr a \cdot \sigma
\end{equation}

for any  $\sigma \neq \varrho_0$ in $\mathfrak{D}$.

Now, we are in position to study exposed points of certain subsets of $\gD$. We begin with
\begin{proposition}
\label{24}
Let $\Tr$ stand for  the trace on a finite matrix algebra. 
 Assume that $a=p-q$ is a
symmetry in $\gD_1\equiv\{\sigma\in\gD:\,||\sigma||\leq 1\}$. Then for any bp matrix
$\sigma\in\gD_1$, $\sigma \neq a$,  one has
\begin{equation}
\label{41}
\Tr a\cdot a > \Tr a\cdot\sigma 
\end{equation}
\end{proposition}

\begin{proof}
Firstly, note that $(f, g) \equiv \Tr f \cdot g$ is a well defined inner product on the self adjoint part of the algebra.
Thus, one has (due to our assumption)
\begin{equation}
\Tr a \cdot  a = \Tr \ I
\end{equation}
and, by Schwarz inequality, putting $\sigma = r - s$ to be a symmetry such that  $\sigma \neq a$,
\begin{equation}
\Tr a \cdot \sigma < (\Tr I)^{\frac{1}{2}} \cdot (
Tr I)^{\frac{1}{2}}  = \Tr I
\end{equation}
Thus, $\Tr a \cdot a > \Tr a \cdot \sigma$.
As a bp matrix  $\varsigma \in {\mathfrak D}_1$ can be written as a convex combination of symmetries $\sigma_i$ (recall that self-adjoint symmetries are extremal points of the self-adjoint part of a unit ball in a \Cs algebra (see \cite{sakai}, or Proposition 7.4.6 in \cite{KR}),
use Corollary \ref{wniosek1} and Corollary 18.5.1 in \cite{Rock}), one has
\begin{equation}
 \Tr a \cdot \varsigma  =  \sum_i \lambda_i \Tr a \cdot \sigma_i <  \Tr I = \Tr a \cdot a,
\end{equation}
where $\lambda_i \geq 0$ and $\sum_i \lambda_i = 1$.
Consequently, $\Tr a \cdot \varsigma < \Tr a \cdot a$ for $\varsigma \neq a$ and the claim follows.

\end{proof}

\begin{remark}
Note that under the assumptions of the above proposition if we put $\varrho_0=a$
then the condition (\ref{41}) has the same form as (\ref{A}).
\end{remark}

As the next step, observe:

\begin{proposition}
\label{15a}
Again, let $\Tr$ stand for the trace on a finite matrix algebra. 
Assume that $a$ is equal to an orthogonal projection
  $p$. 
Then, for bp normalized matrices $\varsigma \in {\mathfrak{D}}_1$ one has
\begin{equation}
\forall_{\varsigma \neq a} \quad \Tr a \cdot a > \Tr a  \cdot \varsigma
\end{equation}
\end{proposition}
\begin{proof}
Again we use the fact that $(f, g) \equiv \Tr f \cdot g$ is a well defined inner product on the self adjoint part of the algebra.
Thus, one has (due to our assumption)
\begin{equation}
\Tr a \cdot a  = \Tr p
\end{equation}
and, by Schwarz inequality, putting $\sigma = r - s$ to be a symmetry (and $\sigma \neq p$)
\begin{equation}
\Tr p \cdot \sigma < (\Tr p)^{\frac{1}{2}} \cdot (\Tr (r-s)p(r-s))^{\frac{1}{2}}  = (\Tr p)^{\frac{1}{2}} \cdot (\Tr p)^{\frac{1}{2}}= \Tr \ p.
\end{equation}
As a general normalized  bp matrix  $\varsigma \in {\mathfrak D}_1$  can be written as a convex combination of symmetries 
one has
\begin{equation}
 \Tr p \cdot \varsigma =  \Tr p \cdot \sum_i \lambda_i \sigma_i = 
 \sum_i \lambda_i \Tr p \cdot \sigma_i <  \sum_i \lambda_i \Tr p  =  \Tr p.
\end{equation}
Thus
\begin{equation}
\Tr a \cdot \varsigma = \Tr p \cdot \varsigma < \Tr p \cdot p = \Tr a \cdot a
\end{equation}
for any $\varsigma \neq p \equiv a$ and
the claim follows (cf. the previous proof).
\end{proof}

A small modification of the above Proposition gives

\begin{proposition}
\label{25}
Again, let $\Tr$ stand for the trace on a finite matrix algebra. 
Assume that $a$ is equal to an orthogonal projection
  $p$, while $ \varrho_0 = p - q$ is an e-symmetry (with $e \neq I$). 
Then, for bp normalized matrices $\varsigma \in {\mathfrak{D}}_1$  and $\varrho_0 \in {\mathfrak D}_1$ (if exist) one has
\begin{equation}
\forall_{\varsigma \neq \varrho_0} \quad \Tr a \varrho_0 > \Tr a \varsigma
\end{equation}
\end{proposition}
\begin{proof}
It is enough to note that $p \cdot \varrho_0 = p$, use $\sigma \neq \varrho_0$, and repeat arguments given in the proof of Proposition \ref{15a}.
\end{proof}

The techniques employed in the proofs of above Propositions lead to

\begin{corollary}
\label{26}
Let us define $\gD_{(n)} \equiv \{ \gamma: n \ \gamma \in \gD \}$ where $n = dim \cH$. Obviously, $\gD_{(n)} \subset B_0(1, \Vert \ \Vert)\equiv \{ a \in \cB(\cH) \otimes_{\pi} \cB(\cH): \Vert a \Vert \leq  1 \} $ (see Corollary \ref{wniosek1}).
Suppose, that $\rho \in \gD_{(n)}$ is an exposed point of $\gD_{(n)}$. Then, $n\cdot \rho$ is an exposed point of $\gD$.
\end{corollary}

Summarizing we have shown that certain projections $P_x$ and bp symmetries are exposed points of $\gD_1$. These results
are especially interesting in views of Basic Lemma which, when augmented by identification of $\varrho_{\varphi}$
 with the transposed Choi matrix (see \cite{St2}) lead to the following conclusions (see also Example \ref{fajnyprzyklad}):
\begin{enumerate}
\item Morphisms correspond to projections of the type: $nP_x$ in $\gD$.
\item Antimorphism correspond to symmetries in $\gD$.
\item Maps of the form $\phi (a) = (f, a f)  I$ where $f \in \cH$, $||f ||=1$ correspond to $|f><f| \otimes I$.
\end{enumerate}

\vskip 1cm

We wish to end this Section with a remark that
 our characterization of $\gD$ indicates that certain $bp$ symmetries could be related with certain non-decomposable maps. To support this claim we note
 that bp condition for a symmetry $s \equiv p - r$ means
\begin{equation}
\label{whooo}
s \quad \text{ is  bp if and only if } \quad   (f \otimes g, s \  f \otimes g) \geq 0
\end{equation}
for any $f,g \in \cH$.
Note that  from (\ref{whooo}) it follows: if $f \otimes g \in p^{\bot}$ then  $f\otimes g \in r^{\bot}$ so $f\otimes g \in (p + r)^{\bot} \equiv 1^{\bot}$.
But this means, compare Definition \ref{defunex}, that the projector $p$ is a projector on a subspace, say $\cH_0$, whose complementary subspace $\cH_0^{\bot}$ contains no simple tensors.  
Conversely, let $p$ be the projector onto the subspace whose complementary subspace contains no simple tensors.
We define a projector $p$ to be simple if
for any simple tensor one has
\begin{equation}
(f \otimes g, p \ f \otimes g) \geq \frac{1}{2} (f\otimes g, f \otimes g)
\end{equation}
where  $f,g \in \cH$.
Suppose that an orthogonal projector $p$ is simple.
Then a symmetry $p - r$ where $r = 1 - p$, is bp symmetry.  

Thus, we got a nice relation with the concept of unextendible product bases. It is worth pointing out that such bases were used by Terhal, \cite{terdal}, in her construction of non-decomposable maps. Here, such maps are appearing in a natural way, as maps ``outside''  the set of completely positive maps.

\section{Convex analysis approach}

In previous section we have seen that the variational approach applied for a description of exposed points is working. However, we are not able to provide a complete list of all exposed points of $\gD$. Moreover, examples presented in the previous Section suggest that extra geometrical aspects should be taken into account.
Therefore, looking for complementary tools, we turn to more analytic approach to this problem with some emphasize on the underlying geometry. We wish to show that there exists an alternative way of characterizing of exposed points of  $\mathfrak{B}^{(+)}_1 ( \equiv  \{  \varrho_{\varphi}: \alpha(\varrho_{\varphi}) \leq 1, \  \varrho_{\varphi}= \varrho_{\varphi}^*, \ \varrho_{\varphi}\geq_{bp} 0 \}$; cf Remark \ref{dobrauwaga}) as well as of $\gD$.  This will be presented in the main result of this Section, in Theorem \ref{wniosek}.

\smallskip

We begin with recalling selected definitions appearing in the convex analysis of real Banach space $X$ (see \cite{Aiz}, Section II.5 in \cite{A}, Chapter 5  in \cite{Megg}, and Chapters 5 and 6 in \cite{Phelps}).

We denote by $S_X$ ($X_1$) the unit sphere (ball) of $X$.
\begin{definition}
A point $x$ of $S_X$ is said to be 
\begin{enumerate}
\item an exposed point of $X_1$ if $\{x\}$ is an exposed face of $X_1$,
\item a rotund point of $X_1$ if every $y \in S_X$ with $||\frac{x+y}{2}|| = 1$ satisfies $x=y$.
\item a smooth point of $X_1$ if there is exactly one element $f$ of $S_{X^*}$ such that $f(x)=1$.
\end{enumerate}
\end{definition}

The sets of rotund points (smooth points) of $X_1$ will be denoted by $rot(X_1)$ ($smo(X_1)$).

If each point of $S_X$ is smooth (rotund) then  the space $X$ is said to be smooth (rotund). The following result says that these two concepts are dual to each other (see \cite{Megg}):
\textit{A reflexive Banach space is rotund (smooth) if and only if its dual space is smooth (rotund).}

Furthermore, it is worth pointing out that smoothness is related to differentiability of the norm (see \cite{Ban}, \cite{Megg}, p. 486)

\begin{theorem}
Let $X$ be a Banach space. $x_0 \in X$ is a smooth point if and only if the norm of X at $x_0$ is Gateaux differentiable.
\end{theorem}

To present an alternative characterization of exposed points of $S_X$ we need one more definition (see \cite{Aiz2}).

\begin{definition}
$x \in S_X$ is defined to be a strongly non-smooth point of $X_1$ if for every $y \in S_{X \setminus \{x\}}$ with $[x,y] \subseteq S_X$, $x$ is not smooth point of  $Y_1$, where $Y=span\{x,y\}$ and $[x,y] = \{ v= \lambda x + (1 - \lambda)y, \quad \lambda \in [0,1] \}$.
\end{definition}

Obviously, $Y_1$ stands for the unit ball in $Y$. The set of all strongly non-smooth points of $X_1$ will be denoted by $nsmo(X_1)$.

We can now quote the promised characterization (see \cite{Aiz2})

\begin{theorem} 
\label{Hiszpan1}
Let $X$ be a real, separable Banach space. Then one has
$$Exp(X_1) = rot(X_1) \cup nsmo(X_1)$$
\end{theorem}

The important point to note here is that to verify conditions appearing in Theorem \ref{Hiszpan1} it is enough to use analytical methods.  
In particular, an analysis of rotundness is based on the following theorem
 (see \cite{Aiz})
\begin{theorem}
\label{Hisz}
Let X be a real Banach space, and $x \in S_X$. Then the following are equivalent
\begin{enumerate}
\item $x$ is a rotund point of $X_1$.
\item for any $y \in S_{X \setminus \{x\}}$,   $lim_{t\rightarrow 0+} \frac{(||x+ty|| -||x||)}{t} < 1$
\end{enumerate}
\end{theorem}

Clearly, the definition of strong non-smoothness is of analytical nature. Thus, both conditions given in Theorem \ref{Hiszpan1} need analytical methods.
However,
observe, that in our case the problem is simplified. Namely, we are interested in an analysis of points of $\mathfrak{D}$. On the other hand, rotundity means (see  \cite{Megg}) that there are no ``room'' for nontrivial straight line segments on the unit sphere. But, $\mathfrak{D}$ is ``flat''!

\vskip 1cm

Let us apply the above results to our problem.
Suppose $\varrho \in Exp(\mathfrak{B}_1)$, where $\mathfrak{B}_1 \equiv \{ \sigma \in \cB(\cH) \otimes_{\alpha} \cB(\cH); \alpha(\sigma) \leq 1 \}$, i.e. $\varrho$ is an exposed point in the unit ball $\mathfrak{B}_1$. If additionally $\sigma \geq_{bp} 0, \quad \Tr \sigma = n$, then $\varrho$ is an exposed point of $\gD$; we have used the ``transition of 
extremality'' discussed in Remark \ref{dobrauwaga}(2). Consequently, to be an exposed point of $\gD$ it is enough that $\varrho$ is  strongly non-smooth, not rotund, is bp positive and finally has the normalization $Tr\varrho = n$.

In particular, using this approach one can verify that the swapping operator, $W$, is an exposed point of $\gD$.

\smallskip

To get the partial converse implication, let us assume that $\sigma$ is an exposed point of $\mathfrak{B}_1^{(+)} \equiv  \{  \varrho_{\varphi}: \alpha(\varrho_{\varphi}) \leq 1, \  \varrho_{\varphi}= \varrho_{\varphi}^*, \ \varrho_{\varphi}\geq_{bp} 0 \}$ such that $\Tr \sigma = n$ ($n = dim \cH$). Thus, $\sigma \in \mathfrak{D}$ and, due to the transitivity, it is an exposed point of $\gD$. We wish to show that 
\begin{claim}
\label{nono}
$\sigma$ is an exposed point of the unit ball $\mathfrak{B}_1$. 
\end{claim}
To this end we need some preparation which will be based on \cite{BaR}, see also \cite{A}, and \cite{WW}.
Let $(X, X^+,||\cdot ||)$ be an ordered Banach space. The norm is $c$-monotone if $0\leq x \leq y$ always implies $||x||\leq c ||y||$ for some positive constant $c$. If $c= 1$
the terminology is simplified, and such norm is said to be monotone.

The cone $X^+$ of $X$ is defined to be $c$-dominating if each $x \in X$ has the decomposition $x =v - w$ with $v,w \in X^+$ and $||v||\leq c ||x||$.
Now we can give (see \cite{BaR})
\begin{theorem}
\label{wow}
Let $(X, X^+, || \cdot ||)$ be an ordered Banach space. Then the following are equivalent
\begin{enumerate}
\item $|| \cdot ||$ is $c$-monotone,
\item $(X^*)^+$ is $c$-dominating,
\end{enumerate}
where $(X^*)^+$ stands for the dual cone in $X^*$.
\end{theorem}

Let us apply the above to our problem.
$X$ is taken to be $\cB(\cH) \otimes_{\pi} \cB(\cH)$. $X^+$ is defined to be the closure (with respect to the norm $\pi$) of the set
$$\{ \sum_i x_i \otimes y_i; x_i \in \cB(\cH)^+, y_i \in \cB(\cH)^+ \} .$$
So, this is the cone defining the order which was used in Basic Lemma.
Further, $X^*$ is equal to $\cB(\cH) \otimes_{\alpha} \cB(\cH)$ while $(X^*)^+$ will be identified with the set of all bp-positive elements in $\cB(\cH) \otimes_{\alpha} \cB(\cH)$.

Now we are in position to continue a proof that the converse implication holds, i.e. to prove Claim \ref{nono}. Our first observation is that the norm $\pi$ is monotone. To see this note: $\sigma \in \cB(\cH) \otimes_{\alpha} \cB(\cH)$ can be written as $\sigma = \sigma^+ - \sigma^-$, $\sigma^{\pm} \geq_{bp}0$, since the set of bp-positive elements form a generating cone and we are considering selfadjoint elements only. Observe that $\alpha (a) \geq ||\Tr_1 a||$ ($\Tr_1$ stands for the partial trace with respect to the first factor). Hence,  it an easy observation that $\alpha(\sigma^+)\vee \alpha(\sigma^-) = ||\Tr_1 \sigma^+|| \vee ||\Tr_1 \sigma^-|| \leq ||\Tr_1 \sigma|| \leq \alpha(\sigma)$. Therefore $\pi$-norm is $1_{+}$-normal and consequently the norm $\pi$ is monotone (for details see \cite{BaR}, pp 226-230).

Further, similar arguments to those given in first paragraphs of Section 3 show that we can take the functional $f$ in Definition 1 to be positive and even to be ``separable ''
i.e. of the form $f(x) = \Tr \varrho_f x \quad$  where $\varrho_f = \sum \Lambda_{ij} P^1_i \otimes P^2_j$ and $\Lambda_{ij}\geq 0$, $P^1_i, P^2_j$ are orthogonal projections.

More precisely, it is enough to combine argument leading to equation (\ref{5a}) with a slight modification of arguments contained in the first paragraphs of Section 3. Thus
$\Tr \varrho_f \sigma > \Tr \varrho_f \sigma^{\prime} \Rightarrow \Tr \varrho_f \sigma + n > \Tr \varrho_f \sigma^{\prime} + \Tr \sigma^{\prime}$, where $\sigma \neq \sigma^{\prime}$, and $\sigma^{\prime} \in \mathfrak{B}_1^{(+)}$. Hence $\Tr (\varrho_f + cI)\sigma > \Tr (\varrho_f + cI) \sigma^{\prime}$
for an arbitrary positive $c$. Therefore, one can always find such positive $c$ that $\varrho_f + I$ is a ``separable''.

To sum up, the (``variational'') condition for $\sigma$ to be exposed point of $\mathfrak{B}_1^{(+)}$ reads
\begin{equation}
\label{53}
\Tr \varrho_f \sigma > \Tr \varrho_f b_0
\end{equation}
where $b_0\neq \sigma$, and $b_0 \in \mathfrak{B}_1^{(+)}$ with $\varrho_f$ a positive, ``separable'' operator.

On the other hand, $\sigma $ will be an exposed point of $\mathfrak{B}_1$ if 

\begin{equation}
\label{54}
\Tr \varrho_f \sigma > \Tr \varrho_f b
\end{equation}
where $b\neq \sigma$, and $b \in \mathfrak{B}_1$. But, as the cone of all bp-positive elements in $\cB(\cH) \otimes_{\alpha} \cB(\cH)$ is dominating (cf Theorem \ref{wow})
$b$ in condition (\ref{54}) has the decomposition $b = b_1 - b_2$ with $b_1,b_2$ being bp-positive, and $|| b_1|| \leq 1$.
Hence (\ref{54}) is equivalent to
\begin{equation}
\Tr \varrho_f \sigma > \Tr \varrho_f b_1 - \Tr \varrho_f b_2
\end{equation}
with $b_1 \in \mathfrak{B}_1^{(+)}$ and $\Tr \varrho_f b_2\geq0$.
But this means that the condition (\ref{53}) implies the condition (\ref{54}). Consequently, the proof of the claim is complete and we arrived at
\begin{theorem}
\label{wniosek}
An exposed point of $\mathfrak{B}^{(+)}_1$ is also an exposed point of $\mathfrak{B}_1$.

Conversely, a bp positive, strongly non-smooth, no rotund $\varrho \in \mathfrak{B}_1$ is an exposed point of $\mathfrak{B}^{(+)}_1$.
\end{theorem}

and

\begin{corollary}
\label{wniosek2}
Strongly non-smooth, no rotund , bp-positive points $\sigma$ of unit sphere (with respect to the norm $\alpha$) having the normalization $\Tr \sigma = n$ are exposed points of $\gD$.
\end{corollary}

To comment on these results we recall that for any linear positive map $\phi$ from a $C^*$-algebra $\mathfrak A$ to $C^*$-algebra $\mathfrak B$ one has 
$||\phi||= || \phi(1)||$
(cf 
\cite{KR}, Lemma 8.2.2.) Further, note that a ``density matrix'' $\sigma$ in $\mathfrak{B}_0 = \{ \sigma \in \mathfrak{B}^{(+)}_1; \alpha(\sigma) = 1 \}$
corresponds to a linear positive map of norm one (cf Lemma \ref{pierwszy lemat}). Therefore, Theorem \ref{wniosek} gives \textit{ the full characterization} of exposed positive linear maps of norm one - exposed both for the set $\mathfrak{B}_1$ as well as for $\mathfrak{B}^{(+)}_1$.
By an argument based on ``transmission of extremality'' (cf Remark \ref{dobrauwaga}(2)), every exposed point of $\mathfrak{B}^{(+)}_1$ (corresponding to a linear positive map of norm one) and belonging to $\gD$ (so corresponding also to unital map) is an exposed point of $\gD$ - see Corollary \ref{wniosek2}.

However, in general, an exposed point of a face $C^{\prime}$, $C^{\prime} \subset C$ maybe not exposed in $C$ (contrary to case of extreme points - see Remark 2.4.4 and the discussion prior to Proposition 2.3.7 in \cite{HUL}).
Consequently, we can not say that every exposed ``density matrix'' $\sigma$ in $\gD$ is also exposed in $\mathfrak{B}^{(+)}_1$.

Finally, it is worth pointing out that every separable Banach space admits an equivalent Gateaux smooth renorming but not with equivalent geometry! Combining this with Lemma \ref{pierwszy lemat}, our results
demonstrate rather strikingly that the geometry determined by the norm $\alpha$ is \textit{the proper, unique choice}.

\section{Final remarks}
\begin{enumerate}
\item
In the presented characterization of normalized bp density matrices $\gD$, so also in the description of positive normalized maps, exposed faces, exposed points,
certain projections and symmetries played  crucial role. However, it is to be expected. Namely, these concepts proved to be very useful in the analysis of the question: which compact convex sets can arise as the state space of unital \Cs  or $W^*$ algebras (see \cite{ASI} and \cite{ASII}). It is worth pointing out that in ``physical'' terms the answer to this question clarifies the statement that Schr{\"o}dinger and Heisenberg pictures are fully equivalent. 
In other words, those  concepts are in the heart of mathematical foundations of algebraic formalism of Quantum Theory. On the other hand, \textbf{positivity and normalization} of maps are necessary elements for the proper definition of quantum probability within Schr{\"o}dinger and Heisenberg pictures. 
Moreover, the relevance of such approach to a description of positive maps, for low dimensional case, was indicated in \cite{Majprep}.

\item The next important point to note here is the possibility of reformulation, now in terms of $\gD$, of the characterization of facial structures of positive maps for low dimensional case which was given recently (see the survey paper \cite{Kye} and references there).

\item
In Section 4 we have seen the strong relation between the characterization of $\gD$ and the geometry of ordered Banach spaces.
It is worth pointing out that there is apparently stronger relation between exposed points and differentiability. Namely, one can extract from \cite{Phelps}, Chapters 5 and 6,  the theorem:  
\textit{The norm $|| \cdot||$ of the ordered Banach space is Gateaux differentiable if and only if its Gateaux derivative is an exposed point (in the weak sense) of the unit ball of the dual space}. However, this leads to differentiation of the projective norm which seems to be rather difficult task.
\item Note, Basic Lemma says that $\mathfrak{B}_1^{(+)}$ corresponds to positive linear maps of norm smaller or equal to $1$. But, (see Section 4), an analysis of exposed points of $\mathfrak{B}^{(+)}_1$ looks simpler. On the other hand, very recently, a progress in the characterization of exposed positive maps in the cone of positive maps was done (\cite{Marcin}, see also \cite{Marcin2}).
\item
Finally, we would like to emphasize again, that although finite dimensional case was assumed, sometimes, we deliberately used more sophisticated notation - the purpose of that is to indicate a possibility for generalizations.

\end{enumerate}
\section{Acknowledgments}

Author would like to thank Erling St{\o}rmer for his kind interest to this work. He is also grateful to
Andrzej Posiewnik and Tomasz Tylec for valuable remarks and to Marcin Marciniak for discussions and for drawing the author's attention to the reference \cite{FLP}. He also wish to thank the referee who kindly offered nice suggestions to improve the original manuscript.
A partial support of the grant number N N202 208238 of Polish Ministry of Science and Higher Education is gratefully acknowledged.

\end{document}